\documentstyle[multicol,aps,epsfig,psfig,epsf]{revtex}

\begin{document}

\draft

\title{Rheology of cholesteric blue phases}

\author{A. Dupuis$^1$, D. Marenduzzo$^{1,2}$, 
E. Orlandini$^3$, J.M. Yeomans$^1$}

\address{$^1$The Rudolf Peierls Centre for Theoretical Physics, 
1 Keble Road, Oxford OX1 3NP, England\\
$^2$ Mathematics Institute, University of Warwick, Coventry CV4 7AL, UK \\
$^3$  INFM-Dipartimento di Fisica and Sezione INFN, Universita' di Padova, 
Via Marzolo 8, 35131 Padova, Italy}

\tightenlines

\maketitle

\begin{abstract}

Blue phases 
of cholesteric liquid crystals offer a
spectacular example of naturally occurring disclination line
networks. Here we numerically solve the hydrodynamic equations of motion to
investigate the response of three types of blue phases to an imposed
Poiseuille flow.  We show that shear forces bend and twist 
and can unzip the disclination lines. Under gentle forcing
the network opposes the flow and the apparent viscosity is
significantly higher than that of an isotropic liquid. 
With increased forcing we find strong
shear thinning corresponding to the disruption of the defect
network. As the viscosity starts to drop, the imposed flow sets the network 
into motion. Disclinations break-up and re-form with their
neighbours in the flow direction.
This gives rise to oscillations in the time-dependent
measurement of the average stress. 


\end{abstract}

\begin{multicols}{2}

In many cholesteric liquid crystals the transition between the regular
cholesteric phase and the isotropic phase occurs through one or 
more intermediate phases known as ``blue phases'' 
(BPs)~\cite{degennes,chandrasekar,mermin}.
Understanding the structure of the blue phases was a theoretical and 
experimental tour de force clearly summarised in~\cite{mermin}.

Blue phases occur because the locally favoured state is a
double twist (i.e. a rotation of the director field in two directions,
as opposed to the single twist of the cholesteric) which is
globally incompatible with the requirement of continuity~\cite{mermin}. 
This results in the formation of networks of defects that separate 
local regions of double twist with axes aligned in different directions.

The networks can be intricate and lead to structures that are periodic at 
large length scales. 
For example the BPI and BPII phases are 3-dimensional cubic phases where 
the orientational order can be periodic on  the scale of the wavelength of 
visible light. Due to this, they have potential applications in fast light
modulators~\cite{heppke91}, tunable photonic crystals~\cite{photonic}
and lasers \cite{lasers}. The main factor limiting the technological
exploitation of BPs is that they are generally only stable 
over a temperature range $\sim 1$ K. However recently it has
been shown that the addition of polymers can stabilise BPI
over more than $60$ K 
opening perspectives in BP technology. 

Though the static structure of BPs is well accepted
essentially nothing is known of their rheology.
Given the intricacy of the director configurations one might 
expect that elastic and viscous responses to an external stress 
combine to give highly non-Newtonian behaviour under flow.
Cholesterics show a strong dependence of viscosity on stress and,
in particular, a striking permeation mode where the backflow acts to
minimise director distortions~\cite{perm}. In blue phases there is also a 
fully three-dimensional network of defects whose elastic properties
will increase the richness of their rheology.

Therefore this paper aims to investigate the rheology of
cholesteric BPs by numerically solving the hydrodynamic
equations of motion for BPI, BPII and, for comparison,
a simplified metastable network which we will term double twist (DT). 
We consider a Poiseuille flow field. Under gentle forcing we find that the
viscosity is consistently higher than that of an equivalent isotropic
phase but, as the forcing is increased, there is strong shear thinning
corresponding to the progressive rupture of the network.
As the disclination network starts to flow the periodicity of the 
blue phase gives rise to an oscillating state which would leave a 
measurable signature in the time-dependent stress curve.
Elastic networks are of interest in many contexts and we compare our
results to very recent studies of the rheological properties of
disclinations in colloidal suspensions in liquid 
crystals\cite{rheology1}.

The equilibrium properties of BPs are described by a
Landau--de Gennes free energy density, ${\cal F}$, written in terms of a tensor
order parameter $Q_{\alpha\beta}$~\cite{degennes,mermin}. This comprises
a bulk term
\begin{eqnarray}
f_{b}=\frac {A_0}{2} (1 - \frac {\gamma} {3}) Q_{\alpha \beta}^2 - 
          \frac {A_0 \gamma}{3} Q_{\alpha \beta}Q_{\beta
          \zeta}Q_{\zeta \alpha} 
+ \frac {A_0 \gamma}{4} (Q_{\alpha \beta}^2)^2
\label{eqBulkFree}
\end{eqnarray}
and a distortion term \cite{mermin}
\begin{equation}
f_{d} = \frac{K}{2} (\partial_\beta Q_{\alpha \beta})^2
+ \frac{K}{2} (\epsilon_{\alpha \zeta\delta }
\partial_{\zeta}Q_{\delta\beta} + 2q_0Q_{\alpha \beta})^2
\end{equation}
where $K$ is the elastic constant and $p\equiv 2\pi/q_0$ is the pitch of the
cholesteric.
$A_0$ is a constant, $\gamma$ controls the magnitude of order
and $\epsilon_{\alpha \zeta\delta}$ is  the Levi-Civita tensor.

The time evolution of the system is governed by the
equation of motion 
\begin{equation}\label{Qevolution}
D_t{Q}_{\alpha\beta}= \Gamma {H}_{\alpha\beta}
\end{equation} 
where $\Gamma$ is a collective rotational diffusion constant
and $D_t$ is the material derivative for rod-like molecules~\cite{beris}.
The molecular field $H_{\alpha\beta}$ is
explicitly given by
\begin{eqnarray}
H_{\alpha\beta} 
& = & -A_0(1-\gamma/3+4Kq_0^2)Q_{\alpha\beta}\\ \nonumber
& + &A_0\gamma(Q_{\alpha\zeta}Q_{\zeta\beta}-
\delta_{\alpha\beta}Q_{\zeta\delta}^2/3)
-A_0\gamma Q_{\zeta\delta}^2 Q_{\alpha\beta}\\ \nonumber
& - & 2Kq_0 (\epsilon_{\alpha\zeta\delta}
\partial_{\zeta}Q_{\delta\beta}+
\epsilon_{\beta\zeta\delta}
\partial_{\zeta}Q_{\delta\alpha}) \\ \nonumber
&+& 4/3Kq_0\delta_{\alpha\beta}\epsilon_{\nu\zeta\delta}
\partial_{\zeta}Q_{\delta\nu}+K\partial_{\zeta}^2 Q_{\alpha\beta}.
\end{eqnarray}
The fluid velocity, $\vec u$, obeys the usual continuity 
equation and the Navier-Stokes 
equation 
\begin{eqnarray}\label{navierstokes}
\rho(\partial_t+ u_\beta \partial_\beta)
u_\alpha = \partial_\beta (\Pi_{\alpha\beta})+
\eta \partial_\beta(\partial_\alpha
u_\beta + \partial_\beta u_\alpha \\ \nonumber
+(1-3\partial_\rho
P_{0}) \partial_\gamma u_\gamma\delta_{\alpha\beta}).
\end{eqnarray}
The stress tensor $\Pi_{\alpha\beta}$
is
\begin{eqnarray}
\Pi_{\alpha\beta}= &-&P_0 \delta_{\alpha \beta} +2\xi
(Q_{\alpha\beta}+{1\over 3}\delta_{\alpha\beta})Q_{\gamma\epsilon}
H_{\gamma\epsilon}\\\nonumber
&-&\xi H_{\alpha\gamma}(Q_{\gamma\beta}+{1\over
  3}\delta_{\gamma\beta})-\xi (Q_{\alpha\gamma}+{1\over
  3}\delta_{\alpha\gamma})H_{\gamma\beta}\\ \nonumber
&-&\partial_\beta Q_{\gamma\nu} {\delta
{\cal F}\over \delta\partial_\alpha Q_{\gamma\nu}}
+Q_{\alpha \gamma} H_{\gamma \beta} -H_{\alpha
 \gamma}Q_{\gamma \beta} 
\label{BEstress}
\end{eqnarray}
where $\rho$ is the fluid density, $\eta$ is an isotropic
viscosity and $\xi$ is related to the aspect ratio of the molecules.
$P_0$ is a constant in the simulations reported here.  Note that
the order parameter field affects the dynamics of the flow field
through the stress tensor. This is the backflow coupling. Details of the 
equation of motion can be found in~\cite{perm}. To solve these equations
we use a three dimensional lattice Boltzmann algorithm~\cite{note,colin}.

In the first column of Fig. 1 we show the equilibrium configuration of the
three different BPs we shall consider. 
The cylinders are disclination lines between regions of double twist
which are repeated periodically to build up the blue phase structure.
Within the tubes the order parameter attains a deep local minimum,
whose value is slightly flow-dependent.
Figs. IB and IC are the accepted structures of BPI and BPII. 
The configuration shown in IA, 
is the DT structure. It is the simplest 
disclination line network that can be constructed with double twist, 
i.e. a periodic array of doubly twisted cylinders with axes along $z$.
DT is metastable and has not been observed. However its simplicity makes it a 
useful tool to help understand the rheology of the other blue phases.
Local regions of the DT configuration are likely to be found
during a temperature quench while a third observed blue phase BPIII
is thought to comprise a gas of doubly twisted cylinders like the ones 
building up the DT network~\cite{degennes,mermin}.

Each structure was obtained by
relaxing ${\cal F}$ to its minimum 
by solving Eq. \ref{Qevolution} numerically.
Periodic boundary conditions were used and initial configurations
were set according to the approximate solutions in \cite{mermin}.
For more details and an equilibrium phase diagram see
\cite{static}.

To study the rheology of BPs, we sandwiched them between two plates a 
distance $L=p/2$ apart along the $z$-axis. {(Qualitatively similar results
are obtained with larger $L$.)}
Poiseuille flow along $y$ was imposed by a pressure gradient 
$\Delta P$ together with no-slip conditions for the velocity at the
plates. We quantify the strength of the forcing via the 
dimensionless parameter $\tilde{F}=\frac{\Delta P L^2}{\eta c}$, where $c$ 
is the sound velocity. At the boundaries we assume that the director profile 
corresponding to the BP structures does not change under flow,
that is the disclinations are fixed at the boundaries. 
Physically, this is likely since small surface irregularities should
pin the equilibrium director profile. {A similar situation occurs
in permeation \cite{perm}, where 
fixed boundary conditions give results closer to the experiments}. 

Fig. 2 summarises the dependence of the apparent viscosity, 
$\eta_{\rm app}$, on the strength of the forcing and compares the blue 
phase behaviour to 
that of a corresponding Newtonian isotropic fluid.
{$\eta_{\rm app}$ is found by computing the flux of $\vec u$
through a plane perpendicular to $y$ 
and by comparing it with that of a Newtonian fluid 
in the same geometry and under the same body force.} 
For small forcing, $\tilde{F}\stackrel{<}{\sim}$
$10^{-2}$, $\eta_{\rm app}$ increases by a factor of $\sim$ $4$ times 
over that of the isotropic fluid. 
All three phases reach a stationary state in which 
the networks of line defects is deformed by the flow, as shown in
column II of Fig. 1. The deformations consist of first bending, and then 
twisting, the disclination line network without changing its topology.
In BP I in particular we observe that the disclinations twist
either separately or in pairs of different handedness, as predicted 
in~\cite{degennes,chandrasekar}.
 
Backflow is significant and acts to decrease the bending and twisting of the 
network lines. This can be clearly seen by comparing the results of the full
simulations in column II in Fig. 1, to those in column III, obtained in
calculations in which backflow was neglected and the order parameter
was simply advected by the flow. 

The primary flow 
is symmetric to a good approximation for BPI and
BPII but displays an asymmetry for DT. This is due to the different 
orientation of the disclination loop with respect to the flow near the
top and bottom plates, which arises as the two disclinations twist. 
In the case of BPI and BPII the disclinations are
more numerous, so that any asymmetry is washed out by averaging.
In the steady state we find  non-negligible secondary flows along both 
the $x$ and $z$ directions.

For small forcing the viscosities of BPI and BPII do not vary greatly. 
The DT phase, however, shows shear thicknening.
This is because of a spectacular change in the defect structure in which each
line of $2\pi$ disclinations (defects of topological strength 1)
characterising the DT structure opens up to a disclination ring
(defect of topological strength 1/2).
The ring then bends and twists on itself as the flow increases. 
A similar transition, driven by applying an external magnetic field, has 
been studied in a system of colloids dispersed in a nematic 
\cite{ter95}.

We now consider what happens as $\tilde{F}$ is 
increased. Fig. 2 gives evidence of significant shear thinning
for all three structures considered. 
The disclination network is destroyed by the flow and the structure
becomes that of a standard nematic subject to a Poiseuille 
flow \cite{degennes,chandrasekar}.

The high and low viscosity behaviours are separated by a crossover 
region in which the viscosity decreases rapidly. 
This corresponds to an intriguing network dynamics, shown, for the 
doubly twisted structure, in Fig. 3.
The disclination lines in a given cell break up and part of the defect
structure moves with the flow into the neighbouring cell. It then re-forms 
the network by attaching to the portions of the defect left behind.
The process repeats and hence there is a continuous process
of destruction and rebuilding of the textures. The dynamics can be monitored
through the time dependence of the free energy and the stress tensor of the
system, both of which oscillate (Fig. 4), at least up to the $1$ ms
time scale reached in the simulations.

Early experiments have reported high viscosities in BPs \cite{negita}. 
It is also interesting to compare our results with 
those obtained very recently in Refs. \cite{rheology1}
for a disclination line network obtained by dispersing 
spherical colloidal particles in a cholesteric liquid crystal.
These authors measured the apparent viscosity versus shear rate and found a 
shear thinning behaviour that closely resembles that shown in Fig. 2.
Finally we comment on the rheology of cholesterics. Here there is an
even higher viscosity at low shear rate, now because the director field
(rather than a disclination network) does not want to distort.
In both systems there is, with fixed boundary conditions, strong shear
thinning at higher shear rates \cite{perm}. In cholesterics when flow is
perpendicular to the helix a nematic-like viscosity is observed
\cite{leslie}.

In conclusion, we have reported the first numerical investigation 
of the rheology of blue phases in cholesterics.
We applied Poiseuille flow starting
from three topologically different equilibrium defect networks: two
correspond to the accepted structures of BPI and BPII, and the other 
is a periodic array of doubly twisted cylinders. 
Under small forcing, the network opposes the flow giving rise to a 
significant increase in the
apparent viscosity $\eta_{\rm app}$. 
Shear forces cause the disclinations to bend and twist. In the
case of the doubly twisted structure we observe a flow-induced unzipping
of the array of lines into a network of loops which causes 
shear thickening at small forcing. As the forcing is increased,
we find clear evidence of shear thinning. 
Similar behaviour is seen in
rheological experiments on colloidal suspensions in liquid 
crystals. In the crossover region, as $\eta_{\rm app}$ drops, 
we also predict a 
novel oscillatory regime which results from 
the interplay of the periodicity of the 
blue phases with the imposed flow. Here the network continuously
breaks and re-forms as portions of the disclinations in the centre of the
channel move to neighbouring cells and relink with the parts of the
network left behind by the flow.

We have considered a director field pinned on the boundaries at $z=0,p/2$.
Experience with permeative flows \cite{perm} suggests that
if free boundary conditions are used
there should be no increase in $\eta_{\rm app}$,
although the rest of the phenomenology  should be
preserved.

Our algorithm should be applicable to study the rheology of other 
3-dimensional disclination networks in liquid crystals
or in liquid crystalline gels, such as those stabilised by 
colloids~\cite{rheology1}, 
emulsions or polymers. An interesting 
related problem is the switching 
in BP devices~\cite{psbp}.



{This work was funded by  EPSRC GR/R83712/01 and by EC 
IMAGE-IN GRD1-CT-2002-00663. }

\begin{figure}
\centerline{\psfig{figure=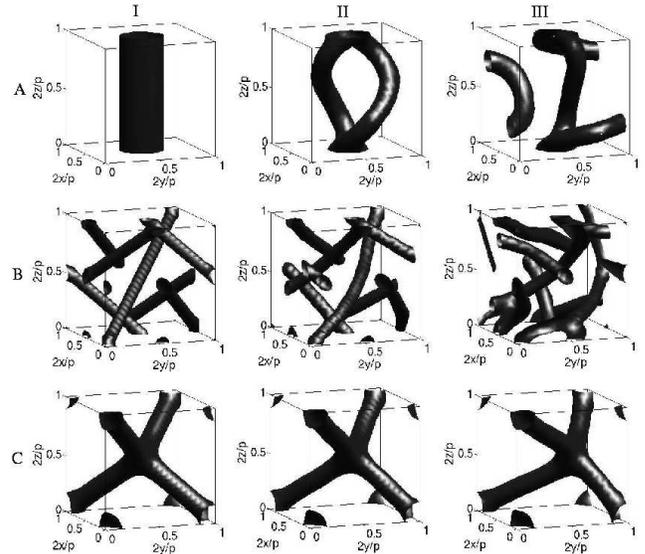,width=8.5cm}}
\narrowtext
\caption{Equilibrium and dynamic steady states for cholesteric BPs.
Each row corresponds to a different BP: A. DT;
B. BPI; C. BPII. Each column corresponds to a different flow
configuration: I. no flow; II. Poiseuille flow with backflow; 
III. Poiseuille flow without backflow. Configurations are drawn 
in terms of their disclination networks, each tube
surrounds a disclination and marks where the order parameter drops
to $60$ \% of its maximum value {(the threshold determines the width of the
tubes, e.g. if it is chosen as $50$ \% the width becomes $\sim$ $10\%$
smaller)}. {The director field is doubly
twisted between disclinations.} 
Parameters were $K=2.5$ $10^{-11}$ N, 
$L=p/2=0.5$ $\mu$m, $\gamma=3$, while $\tilde{F}
\sim 3.3$ $10^{-3}$ for columns II and III;
while the Leslie viscosity ratio $\alpha_3/\alpha_2\sim 0.08$,
so the BP was flow aligning.
BPs are often described by a pair of dimensionless number,
their chirality $\kappa$ and reduced temperature $\tau$ (defined 
in~[3,4]). In our case 
$\kappa=0.59$ and $\tau=0.35$.}
\end{figure}


\begin{figure}
\begin{center}
\epsfig{figure=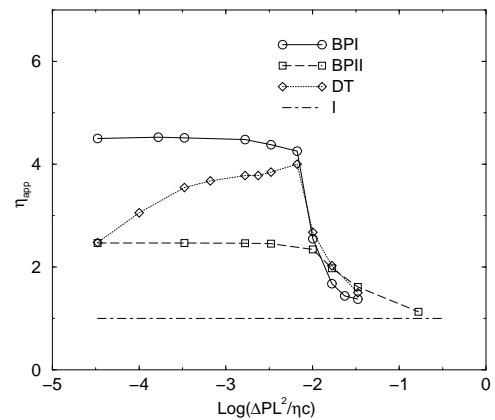,angle=270,width=6.4cm}
\end{center}
\narrowtext
\caption{Apparent viscosities of  BP I,  BPII, DT,
and of a reference isotropic fluid (I),
as a function of  $\tilde{F}=
\frac{\Delta P L^2}{\eta c}$. }
\end{figure}



\begin{figure}
\begin{center}
\begin{tabular}{cc}
1. $6.4$ $\mu$s & 2. $57.6$ $\mu$s \\ 
\epsfig{file=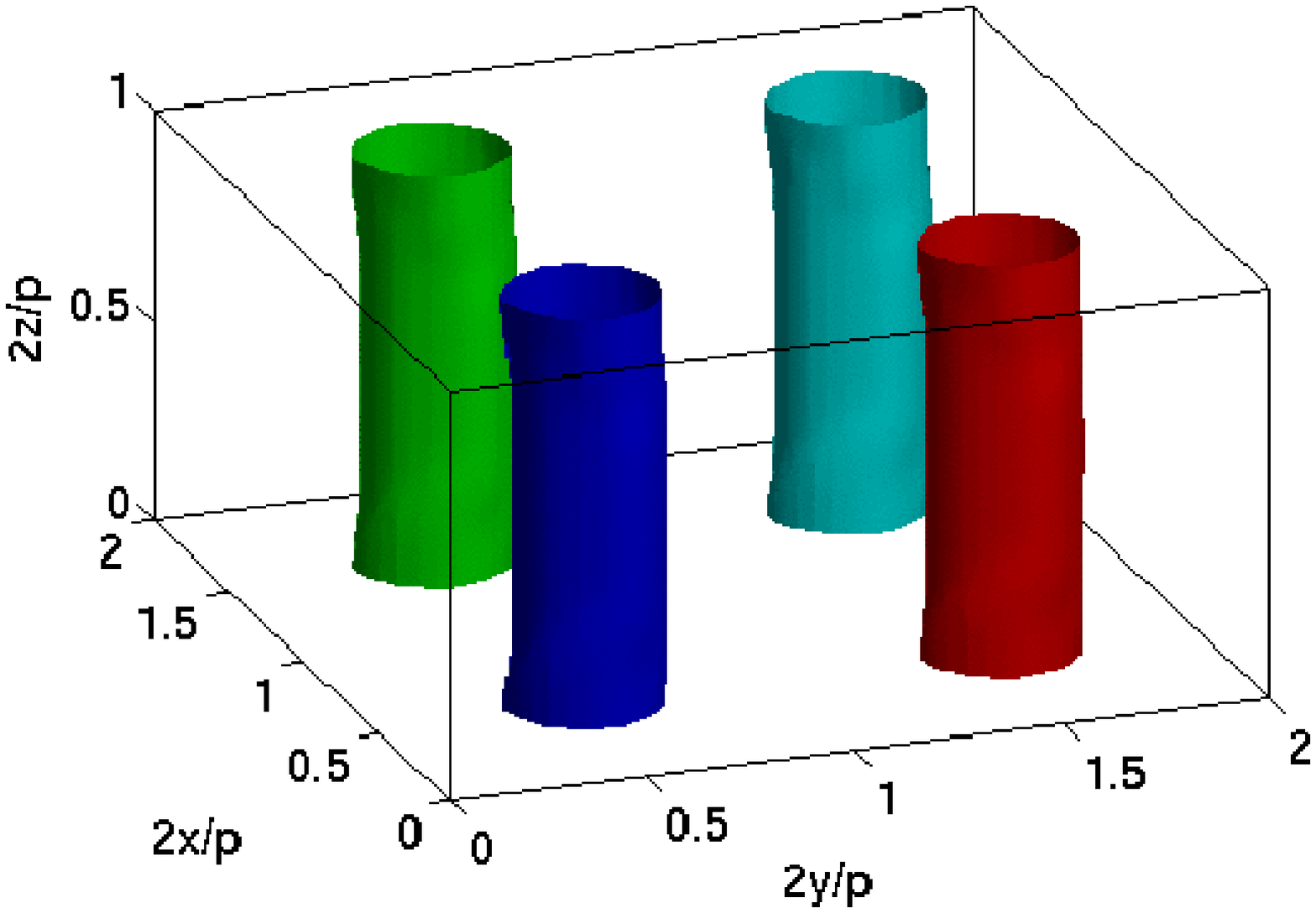,width=4.0cm} &
 \epsfig{file=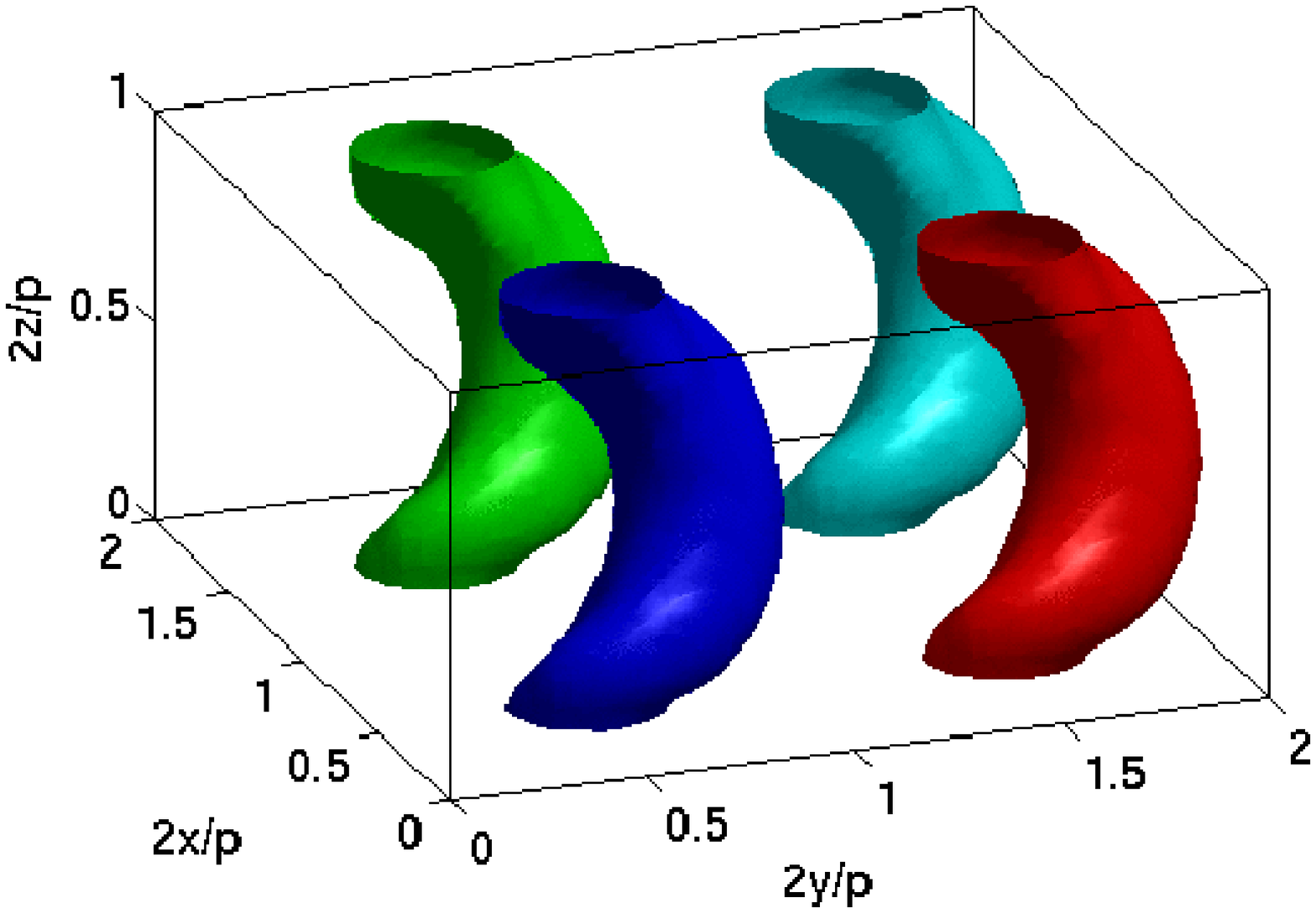,width=4.0cm} \\
3. $0.1$ ms & 4. $0.16$ ms \\
 \epsfig{file=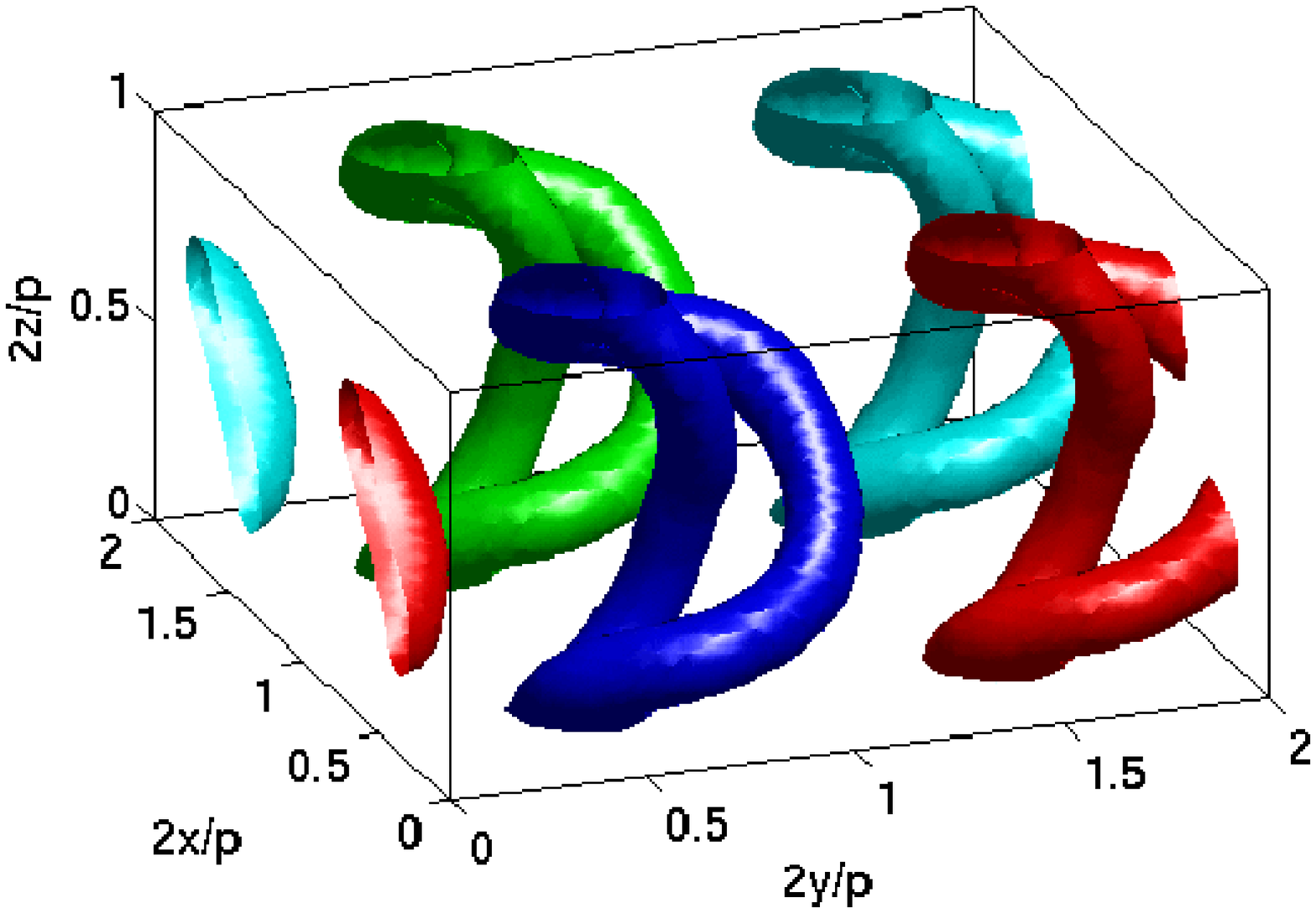,width=4.0cm} &
\epsfig{file=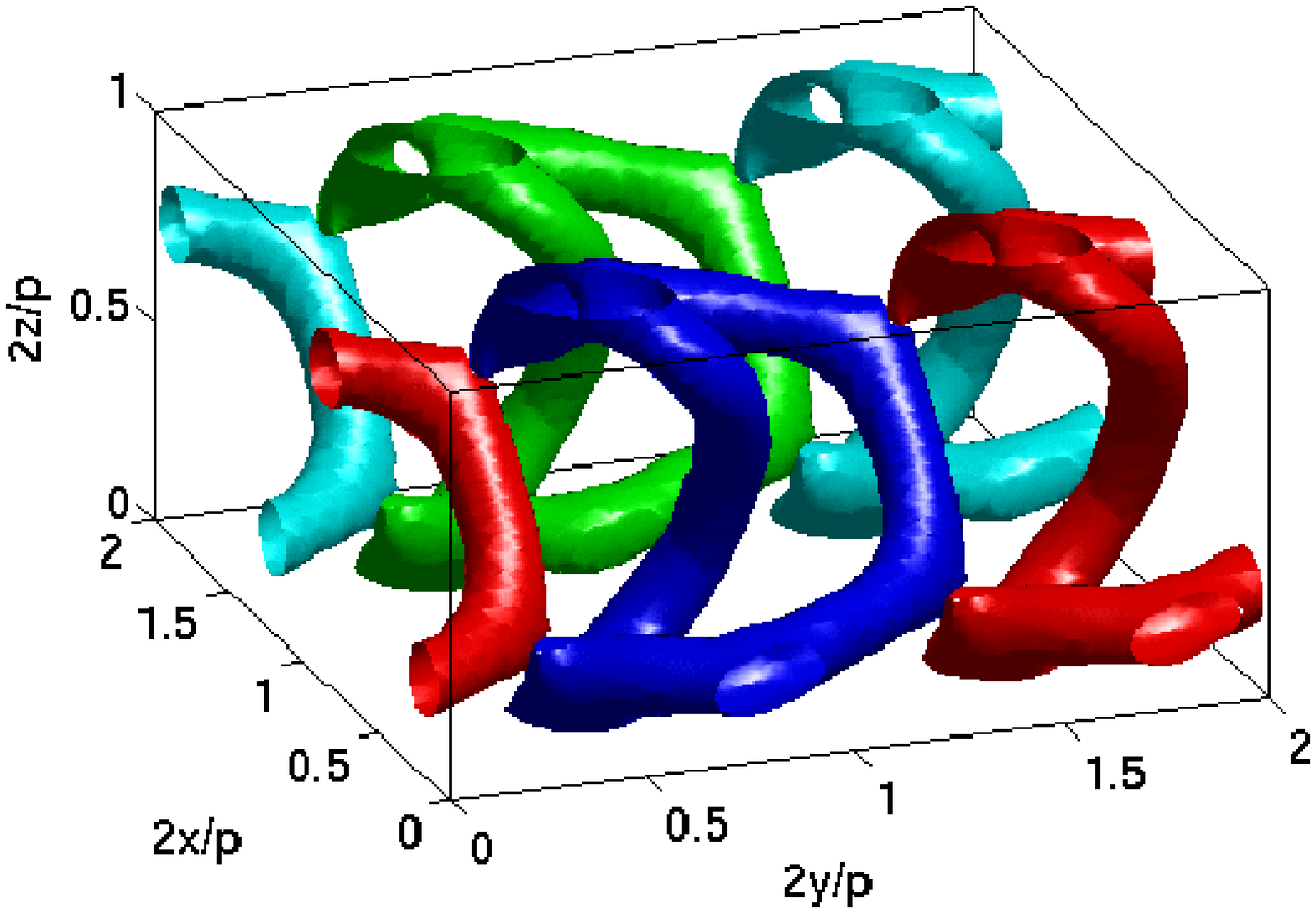,width=4.0cm} \\
 5. $0.2$ ms & 6. $0.25$ ms \\
 \epsfig{file=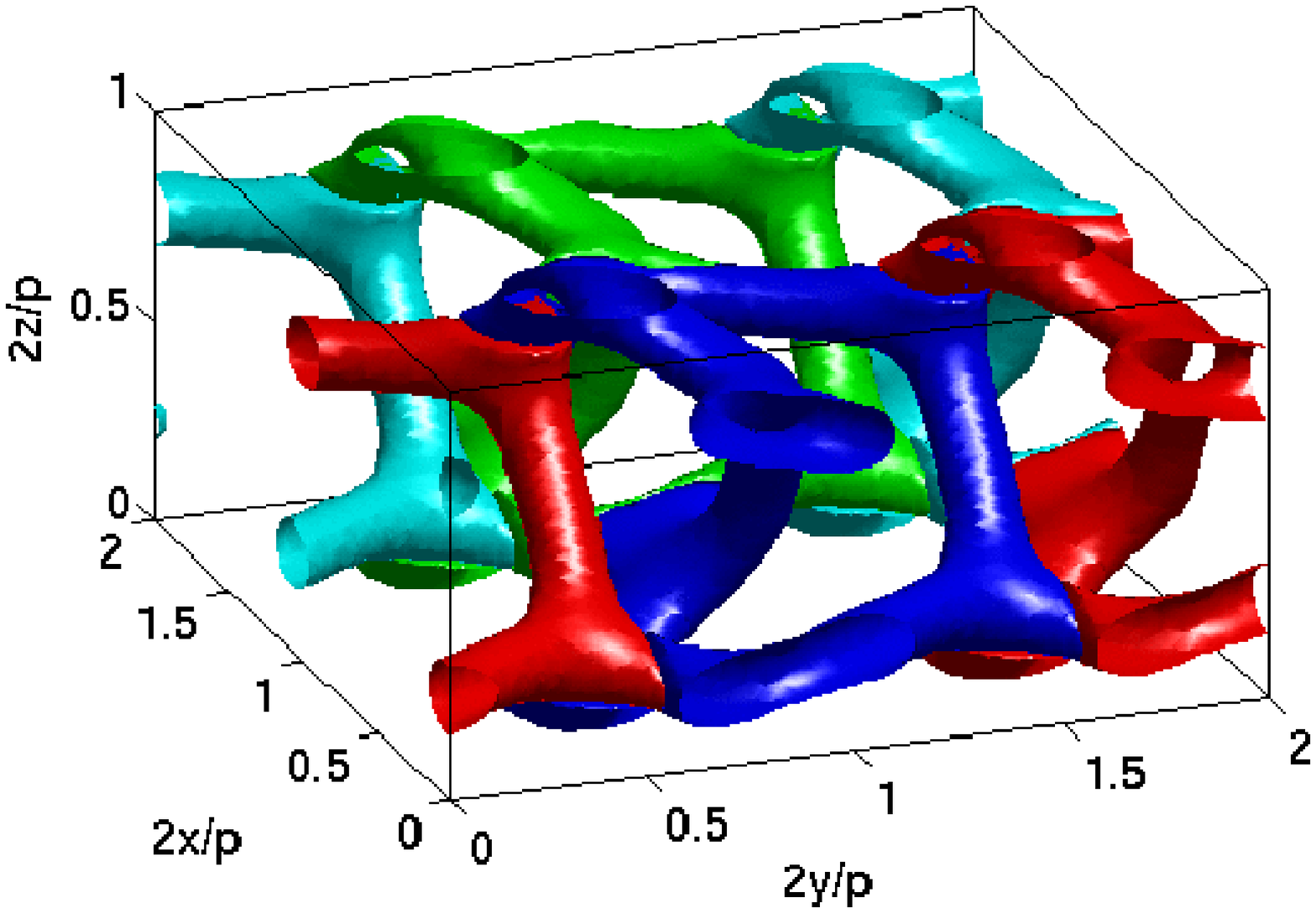,width=4.0cm} &
 \epsfig{file=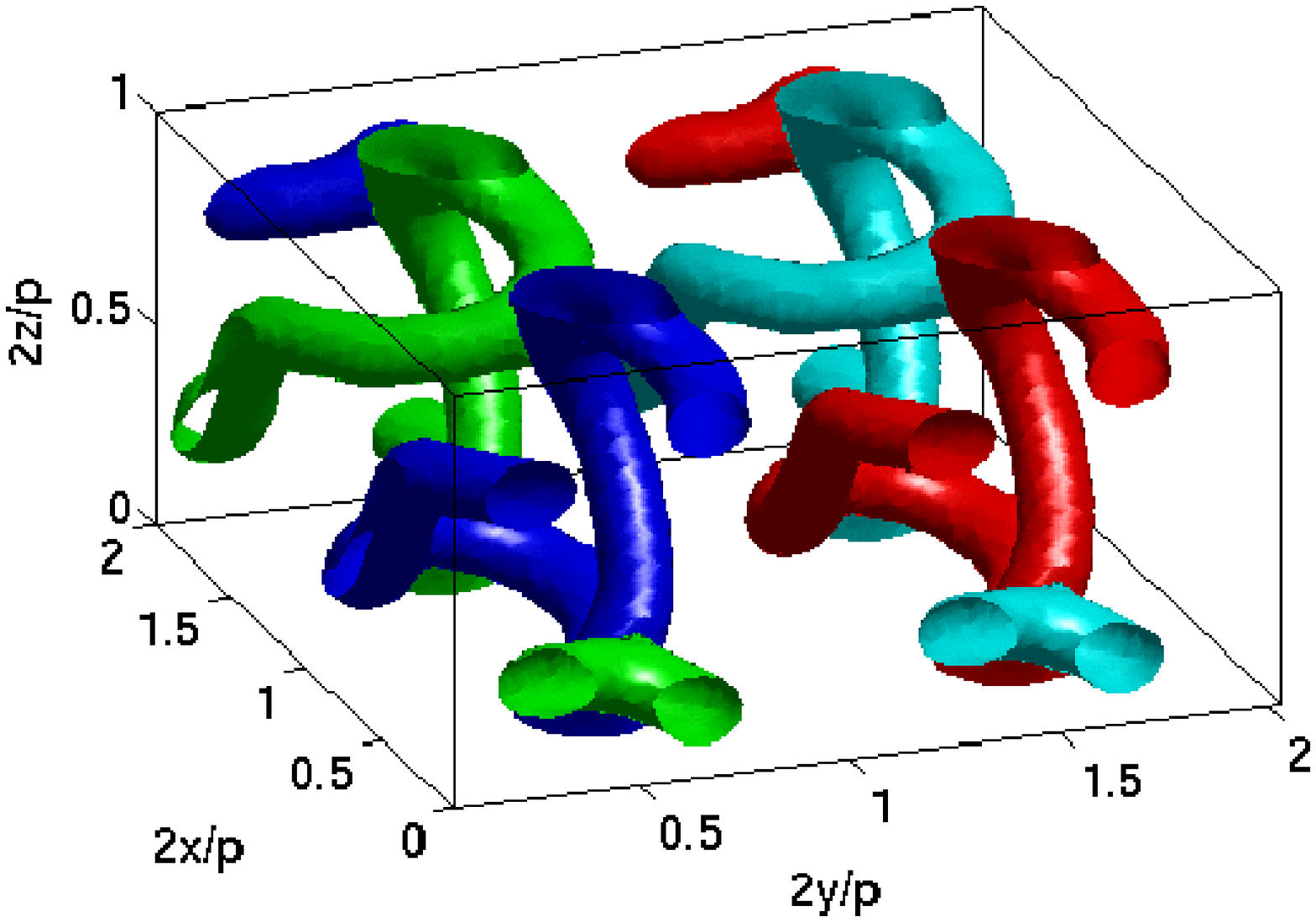,width=4.0cm} \\
7. $0.28$ ms & 8. $0.31$ ms  \\
 \epsfig{file=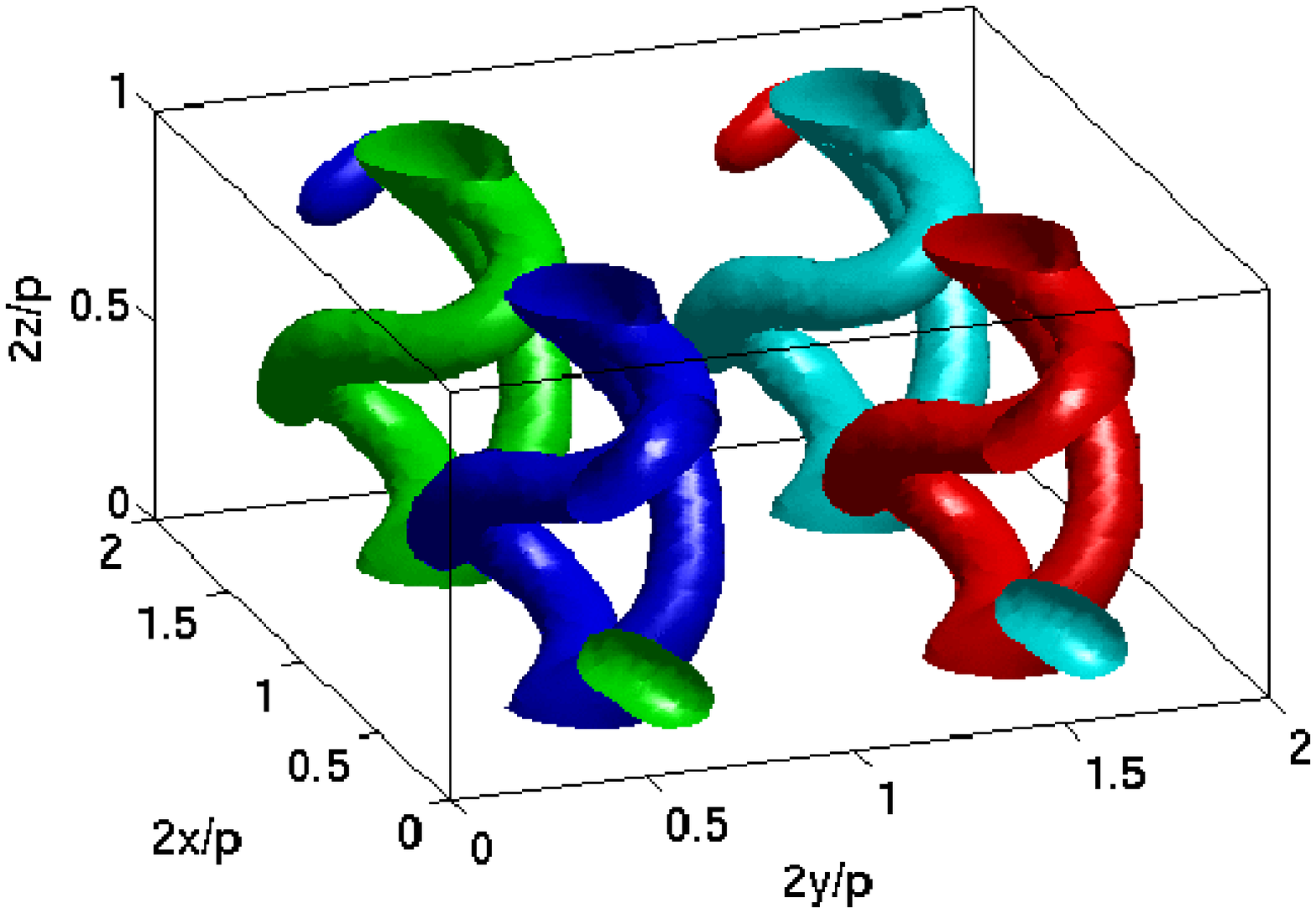,width=4.0cm} &
 \epsfig{file=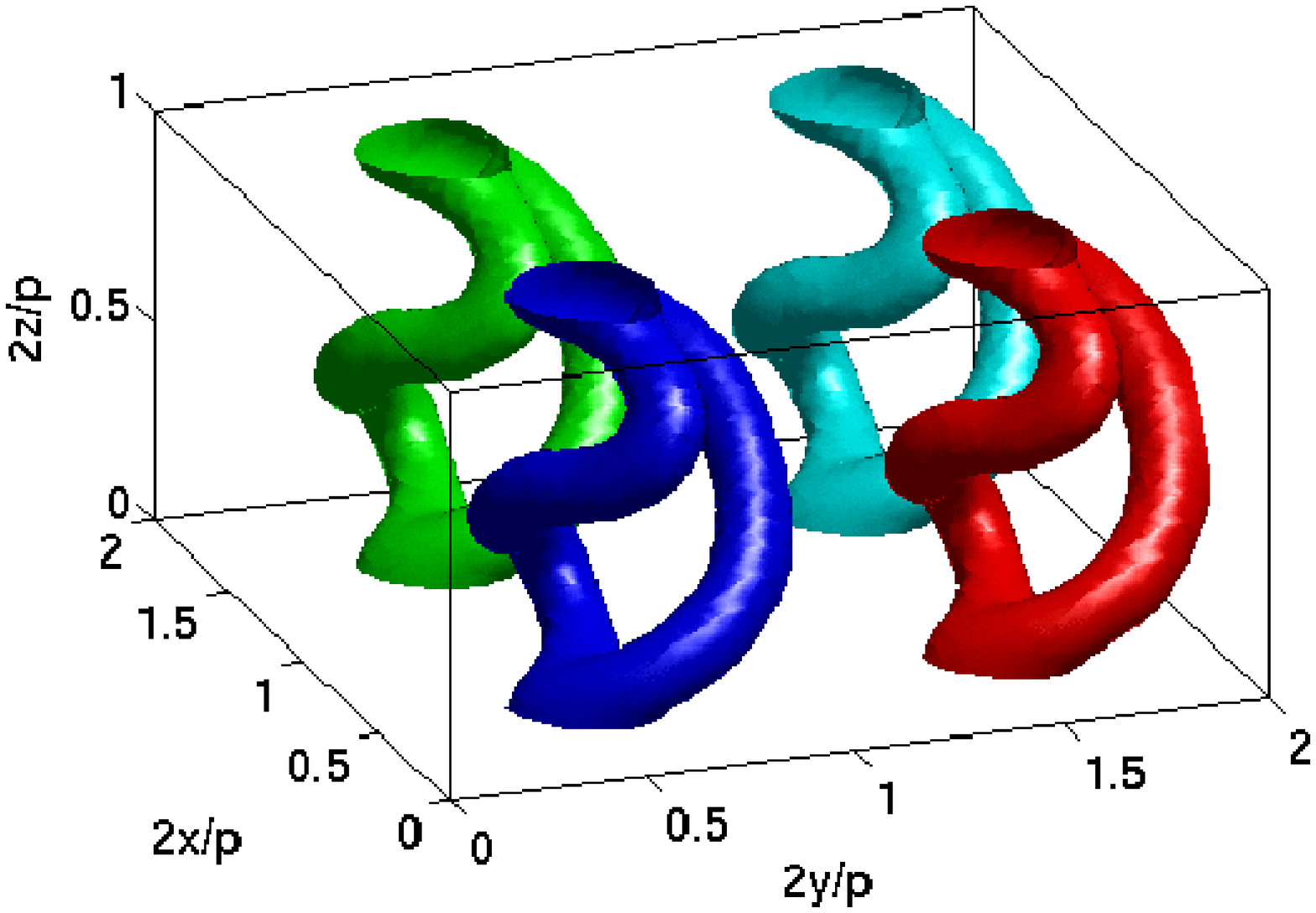,width=4.0cm} \\
\end{tabular}
\centerline{{9. $0.42$ ms}}\\[0.1cm]
 \epsfig{file=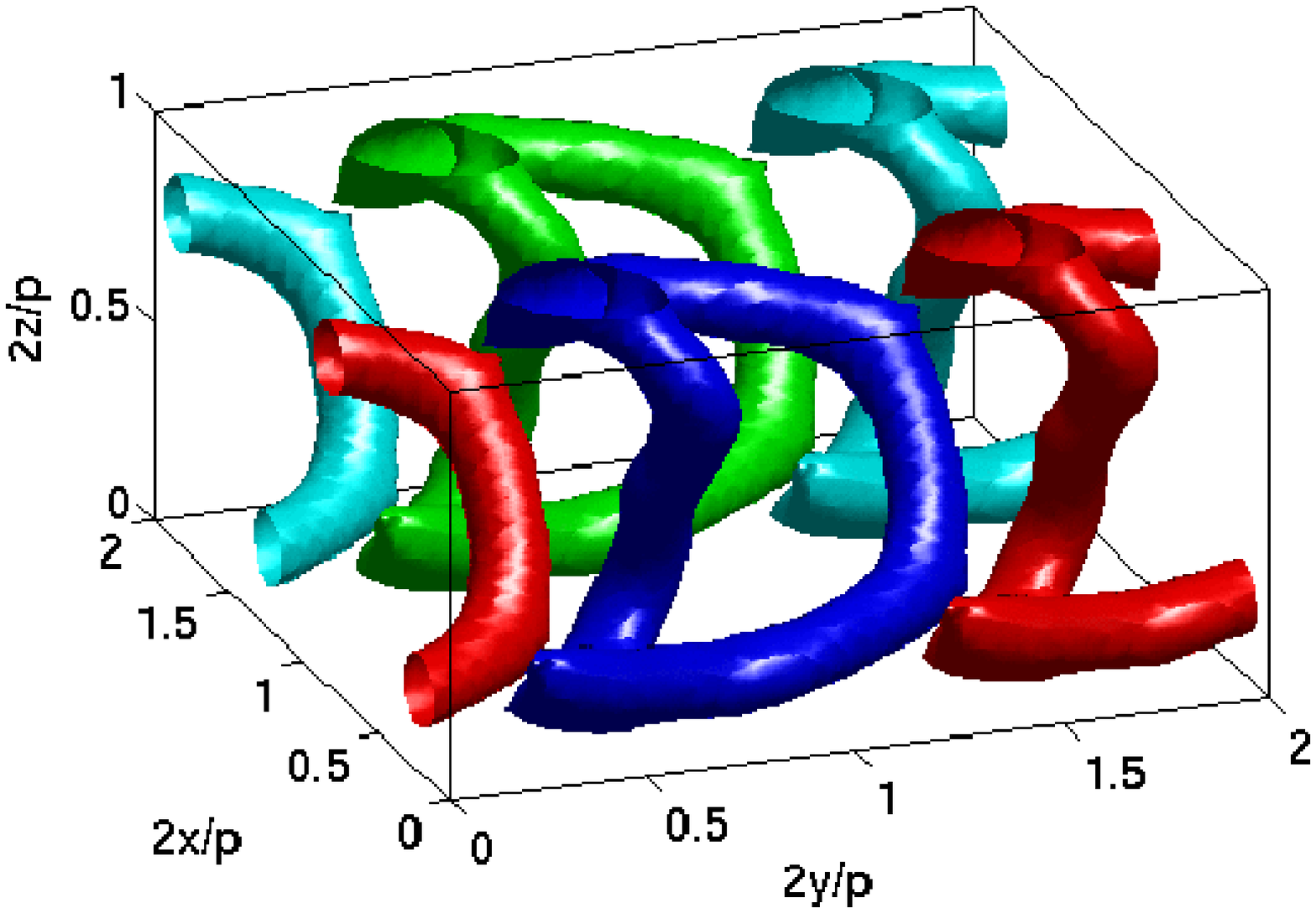,width=4.0cm} 
\end{center}
\narrowtext
\caption{Time evolution of a DT structure under Poiseuille flow,
for $\tilde{F}=0.023$, for which the system
 is in the crossover regime, see Fig. 2. $2 \times 2\times 1$ 
structural periods are shown and distinguished by different colours
{(the actual structure is infinite along both $x$ and $y$)}.
 The disclinations first bend (frame 2) and then unzip (3). 
The unzipped structure near $z=L/2$ is then extended 
in the direction of the flow (4) and merges with its neighbour
along $y$ (5). There is some relaxation (6)  
and then the centres of the disclination
are again stretched by the flow field (7,8) until the structure (9)
resembles (4) and the process repeats. 
Parameters are as in Fig. 1.
The times corresponding to each of the frame are also shown.}
\label{fig:3}
\end{figure}

\begin{figure}
\begin{center}
\epsfig{file=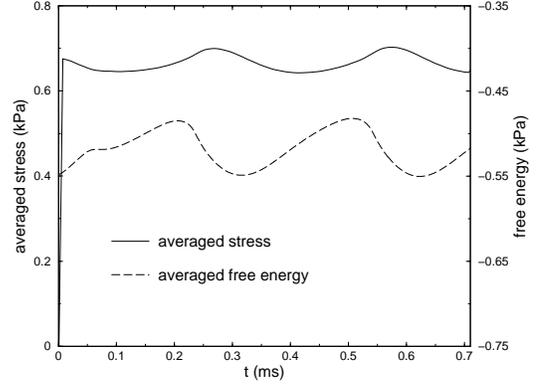,angle=270,width=7.cm}
\end{center}
\narrowtext
\caption{Averaged stress (solid line) and
free energy density (dashed line) corresponding 
to the dynamic evolution shown in Fig. 3. Oscillations are
apparent in both curves. }
\label{fig:sfe}
\end{figure}


\end{multicols}


\begin{thebibliography}{99}
\bibitem{note} Calculations were performed on parallel machines.
A unit cell was discretized into $32^3$ lattice points. In a few cases we 
checked robustness to mesh refinement. On 
average, $10^5$ time steps 
were sufficient to get equilibration and required 
$20$ hours on 8 processors. These correspond to 
$\sim 0.7$ ms. Similar time scales are found 
in \cite{psbp}.  {The algorithm has been validated against 
previously characterised nematic and cholesteric
flows in \cite{perm,colin}.}
\end{thebibliography}
\end{document}